\documentstyle[12pt]{article}
\begin{document}
\begin{center}
{\bf The INT at Age Ten}

\vspace{.45in}

Wick Haxton

{\it Institute for Nuclear Theory, Box 351550\\
University of Washington\\
Seattle, Washington  98195}

\vspace{.25in}

{\bf Abstract}

\end{center}

The history of the Department of Energy's Institute for Nuclear
Theory, located on the campus of the University of Washington,
is reviewed on the occasion of the INT's tenth anniversary.

\vspace{4.0in}
To appear in Nuclear Physics News.

\newpage

The US Department of Energy's Institute for Nuclear Theory (INT) began
operations on the campus of the University of Washington in March, 1990.  In
the decade following, the INT has played an important role in helping the
field of nuclear physics through a period of rapid change.  With the
relativistic heavy ion programs at CERN and Brookhaven, the precision
electroweak efforts at Jefferson Lab, and rapid evolution of sophisticated
solar neutrino experiments, the field's frontiers have broadened to include
the substructure of the nucleon, the properties of QCD at high temperature and
baryon density, and the nature of physics beyond the standard model.  With
this evolution, the intersections of nuclear physics with sister subfields --
astrophysics, particle physics, condensed matter, and atomic physics -- have
become both richer and more entangled.  Theory in particular has profited,
attracting young phenomenologists who would previously have carried some other
label, and becoming enriched with new intellectual challenges like
effective field theory, the pattern of neutrino masses, and
color superconductivity.  Thus, on this tenth birthday of the INT, it is
appropriate to look back at the role the Institute has played in these 
changing
times, and speculate on how it might continue to help the field in the coming
decade.

\vspace{.15in}
\noindent
\underline{
The History}

\vspace{.15in}

The creation of the INT was the culmination of a process begun by the Koonin
Committee, a subcommittee of the Nuclear Science Advisory Committee (NSAC).
NSAC, which advises the US Department of Energy (DOE) and National Science
Foundation (NSF), asked the Koonin Committee to assess the health of 
nuclear theory
and recommend steps that could be taken to strengthen it, given the coming
challenges of JLab and RHIC.  One of the recommendations included in its 1988
report was ``\dots The creation of one or more nuclear theory centers. Such
centers must be truly national in character, must have a significant
interdisciplinary component, and must be viewed and funded as an important
complement to the strengthened individual programs."  In response to the
subsequent DOE announcement, five collaborations submitted proposals for
establishing the INT, which were reviewed by a DOE-appointed selection
committee that visited the proposed sites.  In its selection of
the Seattle proposal, the committee cited its strong and
coherent management plan, its emphasis on young people and visitor programs,
its focus on interdisciplinary problems involving nuclear physics, and the
strong local commitment.  This commitment included a substantial start-up
grant by the Murdock Charitable Trust and a promise by the University of
Washington to create permanent quarters for the INT within its planned new
building for physics and astronomy.
  
The INT opened in the spring of 1990 with a three-month program on Quarks and
Nuclei.  Ernest Henley, who had been instrumental in bringing the INT to
Seattle, served as Director until September, 1991, and was succeeded by Wick
Haxton, the current Director.

Both the staffing of the INT and visitor program selection were (and are) guided by the
INT's National Advisory Committee (NAC).  The NAC consists of nine members who
broadly represent the nuclear physics community, including areas of
intersection with other physics subfields.  The NAC has traditionally included
several European or Asian members, who provide an international perspective
and help maintain closer ties with sister institutes like those at Trento and
Adelaide.  New members, appointed for three year terms, are nominated by
current NAC members, who seek input from the physics community, including the
Division of Nuclear Physics of the American Physical Society.  Nominees must
be approved by the Director of the Division of Nuclear Physics of the DOE.
The NAC traditionally meets each August to review program and workshop
proposals received from the community, advising the Director on their
timeliness and merit, and providing constructive suggestions for promising but
unsuccessful proposals.  The NAC chooses its own chair.  Steve Koonin, John
Negele, Jim Friar, Berndt M\"uller, and Vijay Pandharipande have served in
this role.  Although NAC members volunteer their time, all 30 invitations
extended in the past decade have been accepted.

After assisting the University of Washington with the appointment
of Haxton as Director in 1991, the NAC turned
to the question of the INT's other permanent faculty.  The goal was to 
create a
small but strong ``in house" group who would have broad interests in nuclear
physics and related fields, could contribute to the vigor of the INT's
visitors program, and would stimulate the postdocs and other young people
working at the INT.  The
NAC's 1992 search for the INT's first Senior Fellow resulted in the
appointment of George Bertsch, previously holder of the Hannah Professorship
at Michigan State University.  In 1994 David Kaplan joined the INT as the
second Senior Fellow.  David had been Associate Professor of physics at the
University of California, San Diego.  The INT faculty hold tenured positions
in the University of Washington  Department of Physics.

This faculty hiring was a component of the five-year start-up plan for the INT
that had been negotiated by the DOE and University of Washington.  It
paralleled similar increases in INT visitor activities, the postdoctoral
program, and the administrative staff.  The INT became complete with the
opening of its permanent quarters in the new Physics/Astronomy building on
August 15, 1994.  One weekend
in the middle of an adaptly named summer program (Applications of Chaos),
the INT moved from its temporary quarters in Henderson Hall.  The new
facilities included 22 offices, a seminar room, and a conference room.  Most
important, the INT's immediate neighbors where now the Physics Department's
nuclear, condensed matter, and particle theory groups.  This greatly increased
interactions among INT visitors and the department's faculty, and led to much
more frequent participation in INT programs by department experimentalists and
non-nuclear theorists.  (The department's nuclear theorists, undeterred by the
four-block walk to Henderson Hall, had been active participants in the INT
from the beginning.)

Much of the vitality of the INT derives from its young people.
Generally the INT hosts two Fellows, young researchers appointed for 
five-year terms
who hold the rank of Research Assistant Professor in the Physics Department.
The original siting committee for the INT
argued strongly for these positions:  they anticipated that the visibility of
the position would enhance the careers of the Fellows, and that the Fellows
would help mentor the INT's postdocs.  Three of the four veterans
of this position already hold tenured university positions.
The current occupants are Paulo Bedaque (formerly a postdoc in the Center for
Theoretical Physics, MIT) and Lev Kaplan (formerly a Harvard junior fellow).

In addition, the INT typically hosts five or six postdocs, about half of whom
are supported by various international fellowships.  The INT's young people
are generally quite active in the visitor programs, and have, on a number of
occasions, taken lead roles in the organization of workshops.  For example,
Lev Kaplan teamed with INT postdoc Thomas Papenbrock to organize the INT's
March, 2000, Workshop on Complex Systems and Quantum Chaos.

The INT faculty endeavor to select postdocs based on research promise and
potential for independence, and give little emphasis to the candidate's
overlap with the local research program.  There is a willingness to consider
relevant talent from particle physics, astrophysics, nonlinear dynamics, and
other subfields with connections to nuclear phyics.  The underlying philosophy
is that promising, independent researchers will become intellectually engaged
in our visitor and local activities, and thus in the challenges facing nuclear
physics.  While there is some risk in this approach, the successes appear to
outweigh the failures.  A large fraction of the INT's former
postdocs have already found assistant professor
and staff positions in strong departments (Colorado, Minnesota,
Stony Brook, Brookhaven, and Los Alamos).

Although the INT's original plan made no provision for the faculty to
supervise local graduate students, it soon became apparent that this had been
an oversight.  At the time of the INT's first renewal -- the supporting DOE
grants cover five year periods -- two inadvertent barriers to local graduate
student involvement in the INT were removed.  The second 5-year grant
permitted the support of graduate students as research assistants and also
made it easier for INT faculty to teach, if they desired.  Graduate students
are now a vital part of the INT research effort.

\vspace{.15in}
\noindent
\underline{Programs, Workshops, and Schools}

\vspace{.15in}

The three annual programs are the center of INT activities.  Typically running
for about three months, a program brings visiting physicists to the INT to
focus on some forefront issue facing the field.  Approximately
one-third of the program proposals submitted by community members
succeed on first submission.  Successful proposers are then asked
to organize their programs, helped and guided by the INT's administrative
staff, which handles matters such as visas, housing, and space and budget
projections.  The staff endeavors to free the organizers from as much clerical
detail as possible, leaving the physicists to focus on matters requiring their
scientific judgment (such as which applications to accept and which seminars
to schedule).

A typical program involves 60-70 visitors (though on occasion the total has
approached or exceeded 100).  The average participant spends 3-4 weeks at the
INT, though each program has a few key participants that remain in residence
throughout the program.  Weekly attendence thus is about 18, close to the 
INT's
capacity.  There is significant particpation by experimentalists, who
typically visit for shorter periods.  Overseas participants account for
30-40\% of the total.  Organizers are asked to encourage
the participation of women physicists and others from underrepresented
groups.

Programs attempt to provide participants with long periods for concentrated
individual research and for collaborative work with others.  The primary
mechanism for stimulating collaborations is the morning seminar, where
newcomers are quickly assimilated into the group and where many discussions
begin.  Small working groups often form to continue discussion into the
afternoon and, not infrequently, after dinner.  Participants have keys
allowing access to the INT and to the physics library 24 hours a day.

The programs are scheduled for spring, summer, and fall, with a two-month
break from mid-December to late February.  Traditionally programs with
strongly interdisciplinary themes are scheduled in the summer, e.g., subjects
such as neutrino physics, effective field theory, and neutron stars.  An
important success of the INT has been its popularity in the general physics
community, with strong participation by astro, particle, atomic, and condensed
matter physicists as organizers and visitors.  Summers appear to be the time
when non-nuclear participants are most easily attracted to the INT.

The INT also organizes frequent shorter workshops and miniworkshops, more
intense activities involving a series of seminars and working group
discussions, and lasting typically from two days to a week.  Some of these are
organized as part of and in conjunction with programs.  For example,
organizers may designate the first week of a program as an introductory
workshop, enlarging the program participation for that week.  Others are
independent of the programs.  Over the last five years, an average of four 
such
``stand alone" workshops have been held each year.

Two highly successful efforts are the JLab/INT and RHIC annual workshops.
Both have a similar goal:  to bring together roughly equal numbers of 
theorists
and experimentalists to focus on specific aspects of the programs at these two
major facilities.  Traditionally there has been an emphasis on young
researchers, both as participants and as workshop organizers.  The JLab/INT
series, now in its seventh year, is jointly supported by Jefferson Laboratory
and the INT, with the location generally alternating between these two sites.
The community is invited to suggest topics (and volunteer as organizers).  The
RHIC Winter Workshops, now in their fifth year, are jointly sponsored by
Brookhaven National Laboratory, Lawrence Berkeley National Laboratory, and the
INT.  The site usually alternates between LBNL and the INT.  A steering
committee, comprised of BNL, LBNL, and INT physicists, is responsible for
selecting topics and recruiting appropriate organizers.  Although originally
envisioned as small and very focused workshops involving 25-30 participants,
it has proven impossible to limit attendance, given the strong interest in the
JLab and RHIC programs.  While some of the workshops have grown to 100 or more
participants, the schedule of talks is controlled so that the emphasis remains
on discussions, rather then formal presentations.

The remaining ``stand alone" workshops are generaly one-time efforts focused
on specific topics.  While proposals can be made to the NAC, often workshops
are approved quickly by the Director, so that the workshop lead time is
reduced.  As the cost of a workshop is typically 5-10\% that of a program,
this is a cost-effective way of addressing specific ``hot topics" arising in
nuclear physics.  Often these workshops are organized in collaboration with
outside institutes or universities.  Recent examples include Harvard's
Institute for Theoretical Atomic and Molecular Physics (ITAMP) [Hyperspherical
Harmonic Methods]; Caltech [Nuclear Physics with Effective Field Theory];
Santa Barbara's Institute for Theoretical Physics (ITP) [Time-Dependent
Density Functional Theory]; and Argonne National Laboratoy [Pion Production
Near Threshold].  In such cases the collaborating institution often provides
the site and logistical support, in addition to partial funding.

In order to make the results of selected workshops available to a wider
audience, the INT entered into an agreement with World Scientific to publish
proceedings.  The eleventh volume of the INT series is now in production.
Another mode of communicating and publishing occurs via the INT's web site.
Workshop transparencies can be scanned into the web site.  An audio option
will soon be added.  Applications to attend workshops and programs are now
also handled online.

In addition to programs and workshops, the INT is involved in several programs
that encourage students to pursue physics and nuclear physics.  These efforts
grew out of discussions between the DOE, INT, and NAC at the time of the DOE's
first five-year review of the INT, and reflected concerns about the declining
interest of US students in physics generally.

One effort began in 1995 under the auspices of the NSF's
program, Research Experiences for Undergraduates.
Undergraduates from colleges across the US, most of whom have completed 
their junior years, are brought to
the University of Washington for a 10-week period of intensive research under
the direction of a faculty mentor.  The program is administered by the INT and
operated jointly with the physics department's various research groups.  The
program is very popular, this year generating about 230 applications for 12
positions, with an acceptance rate of nearly 90\% for first offers.  Now in
its sixth year, the program has served more than 80 students, about half of
whom have chosen to do projects in nuclear physics.  One interesting
demographic is the large number of applications from women: nearly
40\% of our REU students are female.  Approximately 20\% of the
class later returns to graduate school at the University of
Washington, and many others are doing advanced studies elsewhere.  One goal of
the program is to introduce students to nuclear physics in the hope that this
will influence future education and career choices.

A second effort is the National Summer School in Nuclear Physics, now in its
13th year.  Sponsored by the NSF, the school is intended for advanced graduate
students and beginning postdoctoral researchers.  Community members volunteer
to host and organize the school.  There is a steering committee, selected by
the Executive Committee of the Division of Nuclear Physics, American Physical
Society, that is responsible for oversight.

Following an unfortunate cancellation of one school seven years ago,
the steering committee invited the INT to take over the school's
administration, hoping this step would help stabilize the effort.
This administrative help -- secretarial support,
advertising, mailing, and poster production, etc. -- is provided through the
INT's DOE grant.  The INT has also hosted the school twice.  The INT has
pledged to continue its support through at least 2004.  The school attracts
40-50 students each year.  The year 2000 school will be held July 3-14 on the
campus of the University of California, Santa Cruz.

\vspace{.15in}
\noindent
\underline{Outlook at Age Ten}

\vspace{.15in}

The scale of activities at the INT, ten years after its founding, is
remarkably different from that envisioned in the original proposal.  That
proposal described three annual programs attracting approximately ten visiting
scientists each.  Today the INT attracts 350 visitors each year.  The list of
Affiliates -- previous visitors and those wanting to be kept informed of INT
activities -- now numbers nearly 2000.  (Many of these
are younger researchers, an emphasis of the INT that was envisioned
in the original proposal.)  The strength of the participation by physicists 
from
other fields, from overseas, and from experiment was not foreseen.  Nor was
the community's enthusiasm for the INT's topical workshops, many of which are
organized collaboratively with other institutes, laboratories, and
universities.  Finally, the original proposers wondered how long the
community could support three annual programs.  Yet the INT receives an 
average of ten
program proposals each year, the quality of which has risen steadily.  The
INT's most common problem is its inability to accommodate the numbers of
applicants who wish to attend the programs.

The success of the INT -- as well as that of its European counterpart the ECT*
-- likely results in part from a sound ``formula" and in part from changes in
the way nuclear physics is now done.  The INT's emphasis on the intersections
with other fields and on collaboration resonates with the changes occurring in
our field due to JLab, RHIC, and SNO.  These facilities address issues
relevant to nuclear, particle, and astrophysics.  There is much anecdotal
evidence that the INT's efforts to expose this physics to a wider community
has led to a broader appreciation of our field.  Similarly, the theory
questions arising from the new facilities are often far more complex than
those of previous times.  The INT's emphasis on collaboration - among
theorists and with experimentalists - has helped theorists focus their
collective energies on difficult but crucial questions.

The INT has evolved in ten years from an interesting experiment to one of the
institutions most responsible for the renewed optimism in our field.  Its
success derives from the support it has received from the community, from the
program proposers who offer their ideas, from the participants who ``vote 
with their
feet" by coming to Seattle, and from DOE program officers who were
willing to back this experiment with consistent funding.  The
INT will endeavor to respond as the
community's needs continue to evolve, and as our field moves forward.

Readers who would like to learn more about the INT -- from its history to its
current and future programs and workshops -- are invited to visit its web 
site,
http://int.phys.washington.edu.

\end{document}